\documentclass[conf]{new-aiaa} 
\usepackage[utf8]{inputenc}
\usepackage{textcomp}
\usepackage{multicol}
\usepackage{lscape}

\usepackage{cleveref}
\usepackage{algorithm}
\usepackage{algpseudocode}
\usepackage{graphicx}
\usepackage{amsmath}
\usepackage[version=4]{mhchem}
\usepackage{siunitx}
\usepackage{longtable,tabularx}
\usepackage{xcolor} 
\usepackage{wrapfig} 
\usepackage{subcaption} 
\usepackage{array}
\newcolumntype{L}[1]{>{\raggedright\let\newline\\\arraybackslash\hspace{0pt}}m{#1}}
\newcolumntype{C}[1]{>{\centering\let\newline\\\arraybackslash\hspace{0pt}}m{#1}}
\newcolumntype{R}[1]{>{\raggedleft\let\newline\\\arraybackslash\hspace{0pt}}m{#1}}

\usepackage{tikz}
\usetikzlibrary{shapes}

\usepackage{nomencl}
\makenomenclature

\newcommand{\comment}[1]{} 

\title{Bridging the Gap: Applying Assurance Arguments to MIL-HDBK-516C Certification of a Neural Network Control System with ASIF Run Time Assurance Architecture
\footnote{Approved for Public Release. Case Number AFRL-2022-5579.}}

\author{Jonathan Rowanhill\footnote{Principal Scientist} and Ashlie B. Hocking\footnote{Principal Scientist}}
\affil{Dependable Computing, Crozet, VA, 22932}

\author{Aditya Zutshi\footnote{Research Engineer, 421 SW 6th Ave.} }
\affil{Galois, Portland, OR, 97204}
\author{Kerianne L. Hobbs \footnote{Safe Autonomy and Space Lead, Autonomy Capability Team 3, 2241 Avionics Circle, AIAA Senior Member.}}
\affil{Air Force Research Laboratory, Wright-Patterson AFB, OH, 45433}

\begin{document}

\maketitle

\begin{abstract}
Recent advances in artificial intelligence and machine learning may soon yield paradigm- shifting benefits for aerospace systems. However, complexity and possible continued on-line learning makes neural network control systems (NNCS) difficult or impossible to certify under the United States Military Airworthiness Certification Criteria defined in MIL-HDBK-516C. Run time assurance (RTA) is a control system architecture designed to maintain safety properties regardless of whether a primary control system is fully verifiable. This work examines how to satisfy compliance with MIL-HDBK-516C while using active set invariance filtering (ASIF), an advanced form of RTA not envisaged by the 516c committee. ASIF filters the commands from a primary controller, passing on safe commands while optimally modifying unsafe commands to ensure safety with minimal deviation from the desired control action. This work examines leveraging the core theory behind ASIF as assurance argument explaining novel satisfaction of 516C compliance criteria. The result demonstrates how to support compliance of novel technologies with 516C as well as elaborate how such standards might be updated for emerging technologies.
\end{abstract}

\section{Introduction}
\lettrine{R}{ecent} advances in reinforcement learning (RL) have demonstrated neural network control systems (NNCS) with better than human performance in military flight scenarios \cite{pope2021hierarchical}. 
However, one of the biggest factors limiting operational use of NNCS is airworthiness certification. Traditional verification requires hundreds of billions of hours of testing to assure safety \cite{kalra2016driving}, which is intractably time consuming and expensive, presenting a significant barrier to use of NNCS on relevant timelines. The Military Airworthiness Certification Criteria, MIL-HDBK-516C \cite{MILHDBK516C}, implicitly assume that the control function itself (e.g. proportional-integral-derivative (PID) control) can be directly verified against the plant/vehicle as a set of simple linear control functions\cite{dillsaver2017military}. Analytical verification techniques have yet to be matured for many complex controller designs, including adaptive and neural network controllers. Some progress has been made in the direct verification of increasingly complex neural network controllers\cite{sidrane2022overt,henriksen2021deepsplit}, however it remains immature. Emerging satisfiability modulo theory (SMT) and other techniques can operate linearly against many benchmarks, although worst-case performance on arbitrary problems will remain exponential \cite{sidrane2022overt,katz2021generating}. The result is that for many problems, direct analysis of neural network control systems may one day be within reach of direct verification using formal methods techniques; however, an alternative approach is needed to field NNCS near term. 

Meanwhile, advances in run time assurance (RTA) \cite{hobbs2021run} technology present a path to enable rapid and safe introduction of NNCS into operational use. RTA-based control architectures filter an unverified primary controller
input by altering unsafe control inputs to explicitly assure safety. When a primary, performance-driven controller is complex, driven by difficult to verify technologies (e.g. a neural network), or dynamically programmed (e.g. learning during operation), RTA can intervene to assure vehicle properties that might otherwise be intractable or impractically expensive to verify.

What is needed, therefore, is a control architecture that effectively pairs an NNCS with an effective RTA, and a means to certify airworthiness of the result. On the latter point, note that \cite{dillsaver2017military} demonstrates that a reversionary switching (simplex) RTA can be shown to satisfy MIL-HDBK-516C criteria for a simple adaptive controller. 
RTA mechanisms such as simplex \cite{seto1998simplex} 
and active set invariance filtering (ASIF) \cite{gurriet2018online} 
are backed by models and reasoning for how they effect safe control. If this reasoning can be formally captured and applied correctly to verification, it could enhance the potential for verification of safety-critical, complex control.

The contributions of this work are as follows.
\begin{enumerate}
    \item The first development of an NNCS-RTA architecture as an abstracted MIL-HDBK-516C system processing architecture (SPA). 
    \item The first development of an argument-based analysis of an NNCS-RTA to satisfy otherwise difficult to verify MIL-HDBK-516C criteria for the NNCS.
    \item Development and presentation of high-assurance compliance methods based on the developed assurance arguments.
\end{enumerate}

The sections of this work are as follows. Section II introduces the concept of an NNCS-ASIF-RTA architecture, safety control properties, and basic verification of vehicle behavior through control with and without RTA. It then shows how assurance argument can represent the reasoning and evidence used to assure safety with RTA. 
Section III presents the NNCS-ASIF-RTA as an abstract system processing architecture (SPA), the results of an investigation of how the use of the architecture impacts conformance with MIL-HDBK-516C certification criteria from MIL-HDBK-516C sections 14 and 15 and examples of where assurance arguments about the NNCS-ASIF-RTA design, backed by verification evidence, can provide a fitting verification method for those criteria. The paper concludes in Section IV with a consideration for how such assurance arguments for novel technologies might aid in amendment of existing standards.

\section{The NNCS-ASIF-RTA Architecture and its Safety Rationale}
This section describes key concepts in this work: ASIF-RTA of NNCS, and safety reasoning for the architecture presented as assurance arguments. In the absence of comprehensive verification approaches for NNCS, an RTA enables strong verification of plant safety properties on the basis of control input filter behavior rather than direct verification of the NNCS function against the plant's entire state space. While this work will focus on an ASIF-RTA and arguments for how it achieves safety, one might apply the same approach to a reversionary RTA, though the required assurance arguments would be very different.

\subsection{Run Time Assurance of Neural Network Control Systems}
\begin{figure}[ht]
\centering
\includegraphics[width=.75\textwidth]{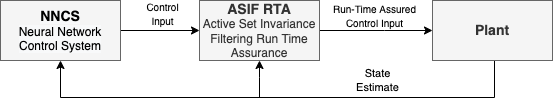}
\caption{A neural network control system paired with an active set invariance filter in closed loop control.}
\label{fig:simpleArch}
\end{figure}

Consider an NNCS, as illustrated in Figure \ref{fig:simpleArch}, that is trained to provide a performance-driven control input to the plant, e.g. an air, sea, space, or ground vehicle. In this design, an NNCS provides control input that is meant for actuation by the plant. However, an ASIF-RTA unit first receives the control input and filters it, modifying the control input if necessary to maintain safety properties of the plant in a verifiable way. Even if the NNCS cannot provide safe control input, the ASIF-RTA assures that input arriving at the plant is safe.

To show compliance with certification criteria, the plant must maintain certain properties, \(V_p\) that are important for safety and airworthiness. Examples of such properties include collision avoidance \cite{hobbs2021formal} and the vehicle remaining within a geofence \cite{hocking2017argument}. 
Without RTA, control signals from the NNCS would always be delivered without modification to the plant. In that case, \(V_p\) is typically assured through inductive argument similar to 
\begin{equation}\label{eq:verSimple}
    C_v \wedge V_v \therefore_I V_p
\end{equation}
where \(C_v\) and \(V_v\) are properties that must be  verified for the controller and the vehicle, respectively, and \(\therefore_I\) is  to be read as `therefore' applied to a conclusion of inductive reasoning. Eq. \ref{eq:verSimple} reads ``controller properties are verified and vehicle properties are verified, therefore, safety and airworthiness properties are satisfied.''
In terms of a MIL-HDBK-516C, \(C_v\) and \(V_v\) are certification criteria of the standard. Intrinsic to the standard is an informal or unrecorded rationale for why the criteria of the standard are sufficient for airworthiness.

When RTA is introduced to assure safety of the NNCS, this changes the required verification properties for the design to the form 
\begin{equation}\label{eq:verWithRTA}
    C_v' \wedge R_v \wedge V_v' \therefore_I V_p
\end{equation}
where potentially more easily verifiable property sets \(C_v'\), \(V_v'\), \(R_v\) are for the controller, vehicle, and RTA, respectively, even if the primary controller is complex, opaque, or dynamically programmed (e.g., an NNCS). Eq. \ref{eq:verWithRTA} reads ``a simpler set of controller properties are verified, RTA properties are verified, and a simpler set of plant properties are verified, therefore, safety and airworthiness properties are satisfied.'' 

The advantage of including RTA is, as stated above, that the required  properties on the controller, vehicle, and RTA might be easier to verify than the set of properties that would otherwise be required on the controller and vehicle without the RTA. There are two resulting challenges.

\begin{enumerate}
\item Existing standards often do not directly align their criterion and verification methods with the target verification properties of the RTA approach. Instead, they often assume that vehicle properties will be assured in a form similar to Eq. \ref{eq:verSimple}, without support for the introduction of RTA and modified verification needs. 
\item The properties to verify control using an RTA are often many and must be very carefully considered. RTAs such as Simplex and ASIF are backed by nuanced design with very particular verification needs. Simplex requires careful consideration of monitoring, backup control design, and switch decision making \cite{hobbs2020elicitation}, while ASIF requires careful determination of resulting control properties under active control signal modification backed by mathematical models \cite{hobbs2021run}. How the required reasoning adds up to the desired vehicle properties must be expressed to the conformance reviewers, otherwise a set of granular and detailed evidence that only indirectly verifies plant safety must be accepted without justification.
\end{enumerate}

In this work, the above issues are addressed by identifying those criterion of MIL-HDBK-516C that assume verification in the form of Eq. \ref{eq:verSimple} and then presenting the reasoning and evidence backing Eq. \ref{eq:verWithRTA}. Assurance arguments are chosen as the format for presenting this reasoning and evidence.

\subsection{Assurance Arguments for Verification of System Properties}

The reasoning and evidence behind verification using RTA (Eq. \ref{eq:verWithRTA}) can be represented as an assurance argument. Assurance arguments are utilized in some safety cultures (e.g. United Kingdom Nuclear Plant Designs, United Kingdom Ministry of Defense Aircraft, Mining Vehicles (Victoria, Australia, EU Airspace, US FDA for Medical Devices) to present a safety case \cite{rinehart2015current}, in which the safety of a system is argued and evidence presented.  Arguments are increasingly used in assurance of novel and emerging technologies, such as in conformance with the UL4600 autonomous vehicle standard \cite{koopman2019safety}. Assurance arguments utilize inductive reasoning, and where possible, deductive reasoning, to show why a claim is true. They can be written as text, but also in modeling languages such as Goal Structuring Notation (GSN) \cite{kelly1999arguing}. 

\begin{figure}[ht]
\centering
\includegraphics[width=1\textwidth]{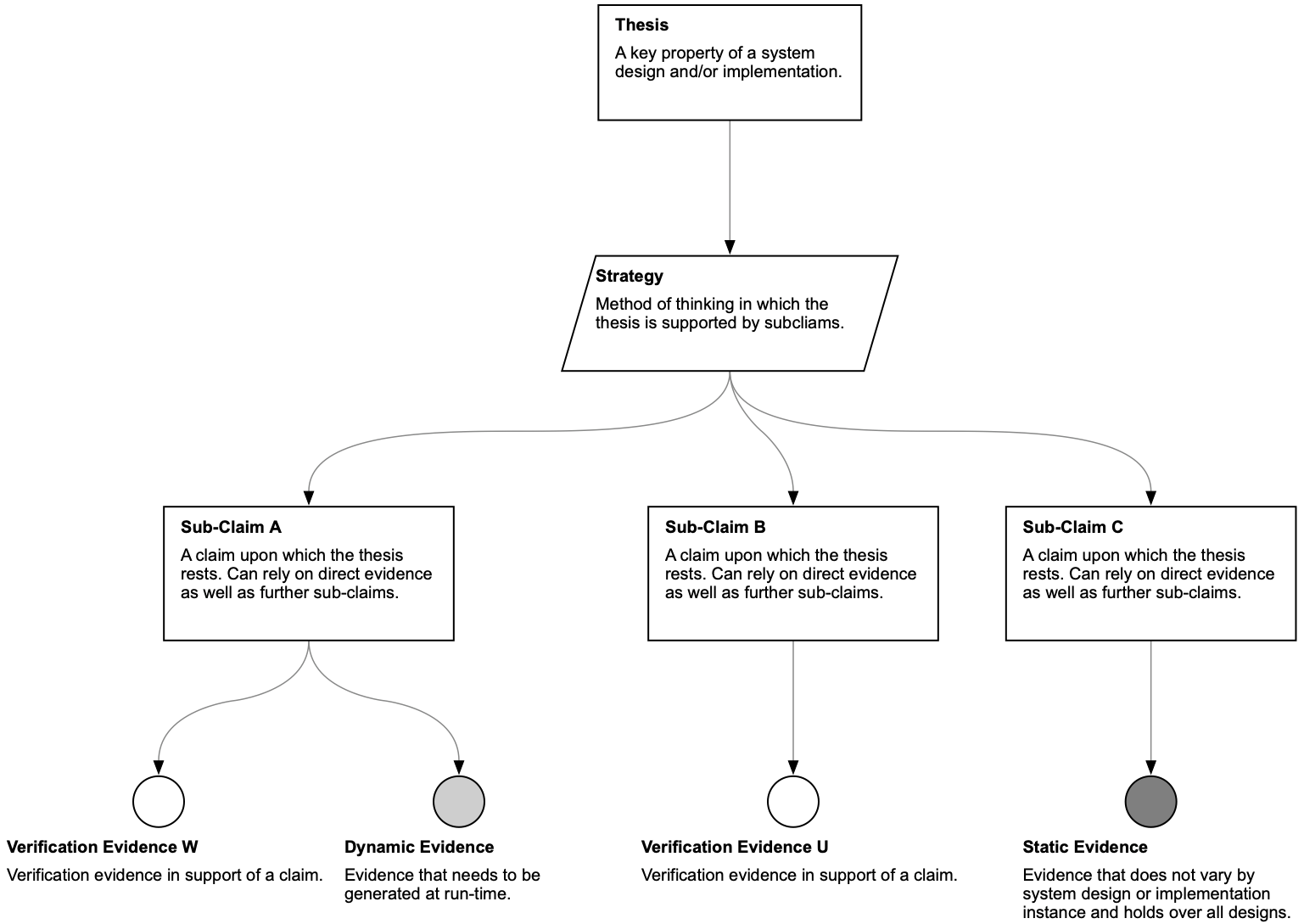}
\caption{A generic assurance argument for a system property expressed in goal structured notation.}
\label{fig:systemTheory}
\end{figure}

An example generic assurance argument for a system property is illustrated in GSN notation in Figure \ref{fig:systemTheory}. In this model, the argument is represented as a tree with the root node (top rectangle) being a \textit{thesis} about a desired property of a system. If this claim could be sufficiently verified by evidence alone (e.g. comprehensive testing), the argument could end with an \textit{evidence} node (circle) attached beneath the root claim, and that would be the end of the argument. But where such direct verification is not possible (as with Eq. \ref{eq:verWithRTA}), further reasoning that is sufficient to verify the root claim is applied. A \textit{strategy} node (parallelogram) describes the method of reasoning over the sum of sub-claims that is sufficient to assure (to the desired level of rigour), the thesis claim. These \textit{sub-claims} (rectangles) can be supported by direct evidence or further strategy and sub-claims. In the figure, they are supported by evidence. This evidence can include static evidence (e.g. textbook knowledge), design time evidence (e.g. testing), run-time evidence (e.g. health monitoring), or other forms of evidence as appropriate.

In the work presented in this paper, assurance arguments are utilized to capture the reasoning behind an RTA design by which evidence collected about the RTA and controller satisfy Eq. \ref{eq:verWithRTA}. Arguments are only used where satisfaction of MIL-HDBK-516C criteria cannot be assured or is not maximally assured using the conventional expectations of listed criterion verification methods.

\section{An NNCS-ASIF-RTA Safety Case}

Three key pieces of work were performed to develop sufficient conformance potential for an NNCS-ASIF-RTA architecture as follows:
\begin{enumerate}
\item \textbf{Abstract NNCS-ASIF-RTA SPA}: An NNCS-ASIF-RTA architecture was developed as an abstract System Processing Architecture (SPA).
\item \textbf{Conformance  Analysis}: The ability of the NNCS-ASIF-RTA to sufficiently satisfy MIL-HDBK-516C conformance analysis criteria of sections 14 and 15 using standard-specified verification methods was analyzed, and where not feasible, assurance arguments were developed as an alternative verification method.
\item \textbf{Assurance Arguments and Methods}: Arguments were developed for selected criteria and the resulting assurance method codified via the required evidence and procedures. 
\end{enumerate}
The remainder of this section discusses these activities and analysis in more detail.

\subsection{Architecture Specification}

\begin{figure}[ht]
\centering
\includegraphics[width=.5\textwidth]{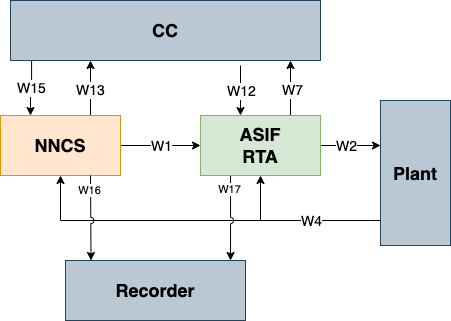}
\caption{An abstract SPA for an NNCS-ASIF RTA subsystem.}
\label{fig:NNCS-RTA-spa}
\end{figure}

An abstract NNCS-ASIF-RTA system processing architecture (SPA) was developed to comply with MIL-HDBK-516C as illustrated in Figure \ref{fig:NNCS-RTA-spa}. The numbering of signals corresponds to a control structure block diagram of the SPA within a larger system documented in \cite{hobbs2023systems}. The design consists of NNCS and ASIF RTA components, discussed previously, that together send control input to the plant. In addition, a command component (CC) configures and operates control modes for NNCS and ASIF RTA. A recorder element provides non-safety critical logging. All components are safety supporting elements (SSEs) except for the recorder. Functions and function threads were identified for each of the components and functional requirements assigned. General vehicle hazards and severity of loss was tied to threads and all threads except recorder monitoring were deemed safety critical.

\subsection{Compliance Planning}

\begin{figure}[ht]
\centering
\includegraphics[width=0.8\textwidth]{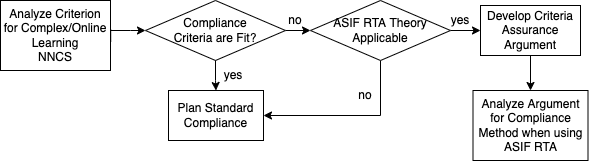}
\caption{Application of argument to strong indirect verification of MIL-HDBK-516C criteria}
\label{fig:method}
\end{figure}

Autonomous control-relevant criteria of MIL-HDBK-516C Sections 14 and 15 were identified and analyzed using the method depicted by the flowchart in Figure \ref{fig:method} to determine where indirect verification through assurance argument was necessary to achieve compliance. First, if the compliance methods stated for the criteria were sufficient and practicable, then no further analysis was performed on the subject criterion. However, if a criterion was deemed not easily satisfied using the stated compliance method for a complex and/or online-learning NNCS, then it was considered a candidate criterion for use of argument. Arguments were developed for the select criterion, and then applied as a specialized method of compliance under the use of ASIF-RTA.

Many criteria from the analyzed sections are relevant to the NNCS-ASIF-RTA control function and would be fulfilled in a compliance effort. In MIL-HDBK-516C Sections 14 and 15, 2 criteria were identified as candidates for indirect verification under ASIF theory: 14.3.3 and 15.2.3.

Table \ref{tab:516IndirectRelevant} presents applicability and limitations of indirect verification for the two criteria recognized as difficult to achieve for a complex NNCS from MIL-HDBK-516C Sections 14 and 15. 
\begin{table}[htb!]
    \centering
    \caption{Criteria Benefiting from Indirect Verification under Strong ASIF RTA Theory}
    \begin{tabular}{m{13em}|m{15em}|m{15em}}
        Criterion & NNCS Conformance Difficulty & Argument Approach Limitations\\
        \hline
        14.3.3: Evaluation of software for elimination of hazardous events
        & Complete verification of coverage of the NNCS control function for elimination of hazards caused by control direction is impractical. & Only applicable to hazards negated by ASIF-enforced safety constraints.\\
        15.2.3: Integration Methodology &  Complete verification coverage of the NNCS control function is impractical. & None\\
    \end{tabular}
    \label{tab:516IndirectRelevant}
\end{table}
Criterion 14.3.3 requires evaluation of software for elimination of contribution to hazards. Argument that can show the correct and failure-free operation of the NNCS-ASIF-RTA control function, where the ASIF-RTA enforces safety constraints that entail prevention of specific hazardous states, assures that the control function does not contribute to those specific hazardous states. Such arguments will
not cover all hazards, but instead those covered
by the ASIF-RTA's safe control output properties. This would partially satisfy the criterion.
Criterion 15.2.3 requires a verification plan for all functions of a developed SPA. This includes verification of functional requirements. A main requirement of the presented SPA's control function is to control the plant so as to continuously satisfy a specified set of safety constraints. Given that ASIF RTA provides strong design reasoning that can be incrementally tested, assurance arguments were developed to capture this reasoning and evidence.
%
Assurance arguments developed for the NNCS-ASIF RTA were fit to the criterion and evaluated for strength and effectiveness. 

\subsection{Development and Application of Arguments}

\begin{figure}[htb!]
\centering
\includegraphics[height=0.9\textheight]{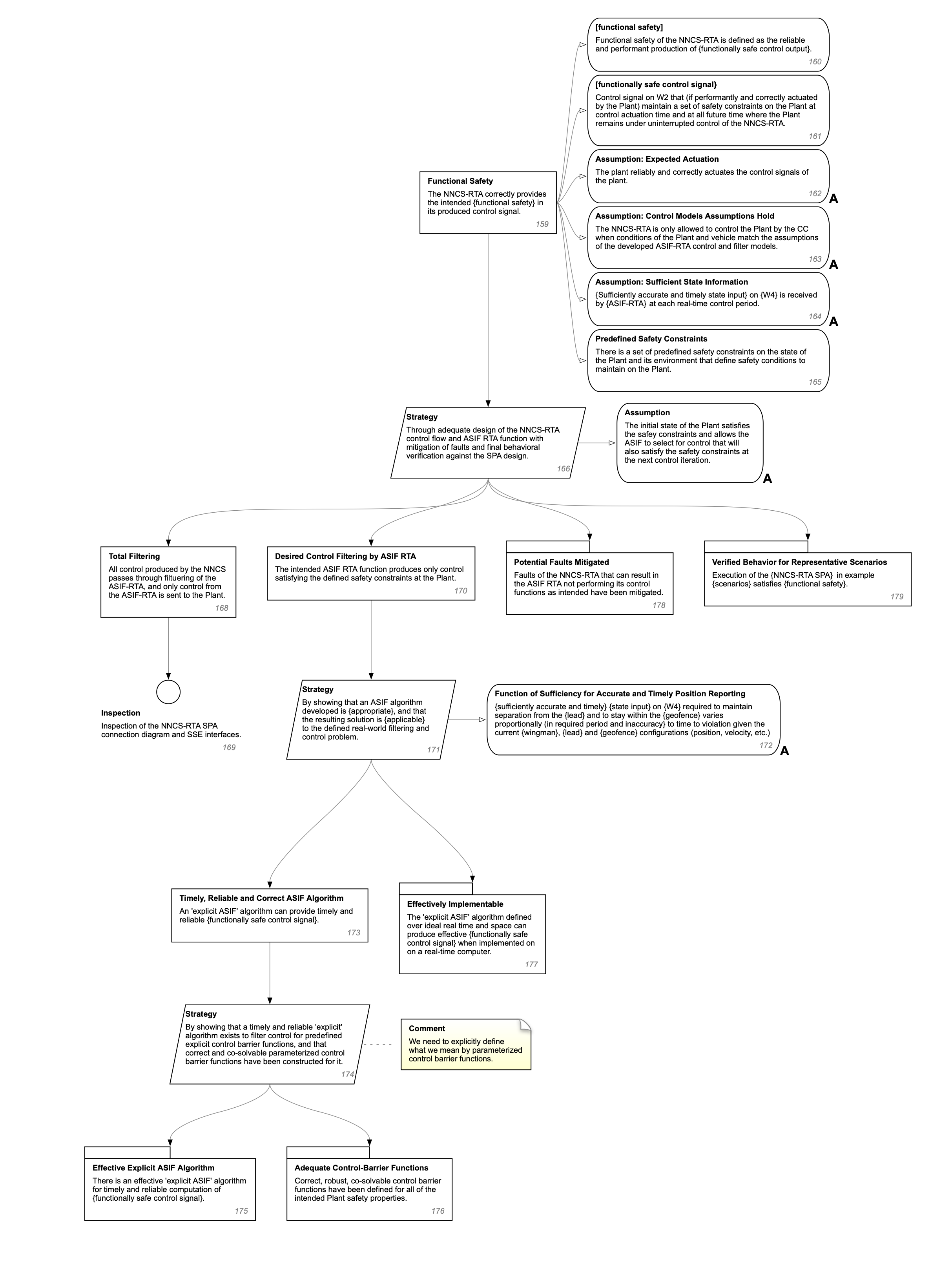}
\caption{The functional safety argument applied to satisfaction of criteria.}
\label{fig:functionalSafetyArg}
\end{figure}

In order to satisfy the above criteria, a functional safety argument was constructed. This argument asserts that the control input generated by the NNCS-ASIF-RTA for the plant satisfies the safety constraints against which the ASIF-RTA was designed. The top-level argument is presented in Figure \ref{fig:functionalSafetyArg}. The root level claim of the argument states: 
\begin{quote}
\textit{Functional Safety of the NNCS-RTA}: The NNCS-RTA correctly provides sufficient assurance that the safety of the NNCS-RTA combined control system, which depends on the RTA component to operate correctly, is assured even when the NNCS outputs an unsafe control input.
\end{quote}
\textit{Functional safety} is defined as the NNCS-ASIF-RTA outputting a \textit{functionally safe control signal}, which is defined as:
\begin{quote}
\textit{Functionally Safe Control Signal from the NNCS-RTA}: A control signal from the RTA to the plant, if performed and correctly actuated by the plant, maintains a set of safety constraints on the plant at control actuation time and at all future time when the plant remains under uninterrupted control of the NNCS-RTA. 
\end{quote}

Support for the above claim is divided between partial direct verification of the claim and reasoning and evidence based on the ASIF-RTA design and its verification. Although direct verification of functional safety cannot be fully assured for a complex NNCS or ASIF-RTA, traditional methods of compliance (simulation, model-based execution, etc.) can show that the SPA functions as intended for representative scenarios and edge cases. Such direct evidence \textit{supports} the argument, but does not sufficiently demonstrate the claim. 

The remainder of assurance is based on ASIF-RTA theory and design, backed by verification evidence. It is argued that the safety constraints hold because all control input from the NNCS is filtered through the ASIF RTA, and the ASIF RTA only outputs control input that satisfies the safety constraints on the plant. This is in turn argued by claiming that there is a timely and reliable algorithm to support the above claim and that it can be implemented effectively. The former claim is satisfied by argument concerning the core theory of an explicit ASIF algorithm that requires development of valid and verified explicit control barrier functions, and that such barrier functions have been developed for the intended safety constraints.

Finally, the entire core argument is strengthened by verifying that faults of an implementation of the SPA are eliminated by specification of the SPA, functional requirements, and explicit safety requirements on the SPA. 
For the purpose of paper scope, argument for failure and fault mitigation, as well as partial direct verification for representative scenarios (and edge cases) are not presented, as the standard methods of verification found in MIL-HDBK-516C are applicable.

\subsubsection{Argument for ASIF RTA}
The argument for an effective explicit ASIF algorithm contends that a reliable and sufficiently performant algorithm exists based on proof in the literature and characterization of the algorithm as a quadratic search problem. In addition, The core induction property of active set invariance asserts that if the present state of the plant satisfies a set of explicit control barrier functions defined for safety constraints, then the ASIF filter can also satisfy those barrier functions in the next iteration. This requires careful construction of barrier functions that assure safety constraints are satisfied and that the next iteration of control will be able to  satisfy them. Papers detailing the mathematics are referred to as evidence.

What remains to be argued is that control barrier functions are constructed for each safety constraint of the plant correctly and robustly. They must also be co-solvable as a group by the explicit ASIF algorithm. Co-solution is evidenced through mathematical analysis.
 
The argument for correct and robust barrier functions is too large to present in a figure. It argues over a specialization of the 'success pattern' \cite{aiello2016model} to show that correct and complete context, definition, and evaluation of control barrier functions has taken place. For example, in arguing sufficient definitions and context, it requires various forms of evidence verifying appropriate vehicle and environment models. It also requires documentation and review showing that control barrier functions are derived from safety constraints and control models correctly, experiments to show that tracking behavior of control is as expected, and demonstration of expected responses, corner cases, and robustness of response under perturbations.

\subsection{Resulting Compliance Method}
Having built the arguments, the method of compliance for criteria 14.3.3 and 15.2.3 can be summarized by the concerns of the argument and the evidence used to support each concern. The method of compliance consists of evidence that must be collected (with an understanding of the backing rationale) in order to satisfy the developed assurance arguments. An example of resulting verification methods is given in Table \ref{tab:correctBarrierEvidence}, showing the evidence that must be collected to assure correct barrier functions.

\begin{table}[!ht]
\centering
\caption{ASIF-Protected Control Compliance Method (Continued)}
\begin{tabular}{b{0.4\textwidth}b{0.4\textwidth}}
   \textbf{Claim} & \textbf{Evidence}  \\
   \hline
   Correct Safety Constraint & Constraint Specification \\
   & Peer Review \\
   & Authority Sign-Off \\
   Sufficient Entity Dynamics Models & Model Analysis and Review \\
   & Higher-Fidelity Simulation and Sim Validity Checks \\
   Sufficient Env. Disturbance Models & Table of Limits (Parameters) \\
   & Standards and Literature Review \\
   & Domain Expert Review \\
   Sufficient Tracking Models & Models Analysis \\
   Known Control States, Rates, and Latencies & Documentation \\
   Correct Mathematical Derivation of Barrier Functions & Mathematical Derivation \\
   & Peer and Domain Expert Review \\
   Acceptable Tracking Error & Simulation Results \\
   & Simulation Validity \\
   Empirical Correct Control Examples & Representative
   Plant Plots \\
   Empirical Verification for Corner Cases &
   Corner Case Analysis \\
   & Peer and Domain Expert Review \\
   & Resulting Plant Plots \\
   Empirical Verification for Off-Nominal Conditions & 
   Model Analysis \\
   & Peer and Expert Review of Conditions \\
   & Resulting Plant Plots \\
\end{tabular}
\label{tab:correctBarrierEvidence}
\end{table}

Overall, the developed assurance arguments require 51 pieces of evidence, each of which can be thought of as part of this specialized verification method for the above criteria. Table \ref{tab:evidence1} shows a categorization of the evidence by type. A significant quantity of evidence involves review and development of analytical models which are tested in validated simulations. Some static analysis is performed on the architecture to assure control signal feeds at the right times. Safety requirement development is assured with a STAMP/STPA method and evidence, and many other categories of evidence are required to assure that assumptions and guarantees of plant, control, and ASIF RTA align.

\begin{table}
\centering
\caption{Evidence and User-Answered Goal Types}
\begin{tabular}{b{0.5\textwidth}b{0.2\textwidth}}
   \textbf{Type} & \textbf{Count} \\
   \hline
   Proof, Equations, and Mathematical Analysis & 11 \\
   Requirements and Assume Guarantee Analysis & 8 \\
   Simulation Input Analysis & 8  \\
   Peer and Expert Review & 6 \\
   Simulation Results & 5\\
   Static Analyses & 3 \\
   Documentation & 3 \\
   Tool Validation & 2 \\
   Model Sufficiency Analyses & 3 \\
   Numerical and Discrete Time Stability Analyses & 2 \\
   STPA Tables & 1 \\
   Computational Cost Analysis & 1 \\
   Performance Analysis and Testing & 1 \\
   Goals Left to Implementer to Satisfy & 1 \\
\end{tabular}
\label{tab:evidence1}
\end{table}   

It would be expected that the resulting evidence be collected for any application of the provided ASIF-NNCS-RTA in order to satisfy criteria 14.3.3 and 15.2.3 using the above arguments. If such evidence is supplied and found satisfactory, then it should be the case that the criteria are satisfied to the extent that the arguments are strong.  In addition, assurance arguments should be constantly updated as flaws or additional concerns are discovered. This can in turn lead to  new required evidence. 

\section{Summary and Conclusions}

This presented work demonstrates how an NNCS safeguarded by RTA for specific safety properties on a controlled plant (e.g. vehicle) can satisfy airworthiness criteria of MIL-HDBK-516C that can be difficult or cost-prohibitive to satisfy using traditional verification methods. The work focused on sections 14 and 15; however, future work could more fully analyze the MIL-HDBK-516C standard's criteria. Limited time prevented analysis of many important sections of the document, for example, sections 4 and 6, related to systems and control behavior. The completed work involved careful selection of an RTA architecture with strong control properties, namely an active set invariance filtering RTA, and development of alternate verification methods for some criteria of the standard. The alternative verification method consisted of assurance arguments, modeled in GSN. The  result is a set of evidence that must be collected in order to satisfy the arguments, which in turn should satisfy the criteria, and therefore represent an alternative verification method. An overview of the arguments and required evidence was presented.

In conclusion, we expect that the approach taken in this work might be valuable as other new and novel architectural innovations are applied to aerial systems, such as for technologies in the autonomous vehicle domain. The quality of the resulting arguments must always be vetted by domain and regulatory experts, but can be part of assessing novel technologies by responsible parties. Furthermore, the resulting alternative verification methods can be rigorous and detailed, and the resulting argument and evidence models can serve as input to future iterations of standards to which they are applied should the covered technology end up becoming commonplace. For less common technologies, such alternative verification methods might be placed in an annex of conformance capabilities. The NNCS-ASIF-RTA MIL-HDBK-516C section 15 compliance case can be viewed at \url{www.dependablecomputing.com/nncs-asif-rta/case.html}.

\section*{Acknowledgments}
The authors would like to thank Matt Dillsaver, John Schierman, David Kapp, Ray Garcia, Natasha Neogi, Mallory Graydon, Michael Holloway, Suresh Kannan, Sean Regisford, Benjamin Heiner, and others for their input and feedback on this work. This work was supported by the Test Resource Management Center and the Air Force Research Laboratory ADIDRUS Contract. The views expressed are those of the authors and do not reflect the official guidance or position of the United States Government, the Department of Defense or of the United States Air Force.


\bibliography{references}

\end{document}